\documentclass{aastex}
\usepackage{spr-astr-addons}
\usepackage{url}
\usepackage{graphicx}

\RequirePackage{color}

\newcommand{\emaila}{istomin@lpi.ru}

\def\Xint#1{\mathchoice
 {\XXint\displaystyle\textstyle{#1}}%
 {\XXint\textstyle\scriptstyle{#1}}%
 {\XXint\scriptstyle\scriptscriptstyle{#1}}%
 {\XXint\scriptscriptstyle\scriptscriptstyle{#1}}%
 \!\int}
 \def\XXint#1#2#3{{\setbox0=\hbox{$#1{#2#3}{\int}$}
 \vcenter{\hbox{$#2#3$}}\kern-.5\wd0}}
 
 \def\dashint{\Xint-}

\begin{document}

\title{Acoustic $\alpha$-disk}
\shorttitle{$\alpha$-disk}
\shortauthors{Istomin}

\author{Ya.~N. Istomin \altaffilmark{1,2}} 


\altaffiltext{1}{P.N.~Lebedev Physical Institute, Leninsky Prospect 53, Moscow 119991, Russia}
\emaila
\altaffiltext{2}{Moscow Institute Physics and Technology, Institutskii per. 9, Dolgoprudnyi, Moscow region, 141700, Russia}

\begin{abstract}
It is shown that the turbulent flow of acoustic waves propagating outward from the inner edge of the disk causes the
accretion of the matter onto the center. The exponential amplification of waves takes place in the resonance region, $ \omega = (n\pm 1)\Omega $.
Here $ \omega $ is the frequency of the acoustic wave, $ n $ is its
azimuthal wave number, $ \Omega (r) $ is the angular frequency of rotation of the disk. The effect is similar to
the inverse Landau damping in a collisionless plasma. Energy comes from the energy of rotation of the disk. That leads
to decrease of the disk angular momentum and to accretion of the matter. The value of the accretion rate $dM/dt$ is ${\dot M} =
\pi rc_s \Sigma_0 (c_s / v_{\phi 0})^2 W $. Here $ c_s $ is the speed of sound of the disk gas, $ v_{\phi 0} $ is the Keplerian rotation velocity,
$\Sigma_0 $ is the surface density of the disk, $ W $ is total power of the acoustic turbulence, 
$ W \simeq \int_0^\infty d \omega \sum_{n\geq 0} \Big {|}
\frac {\Sigma'} {\Sigma_0} \Big {|}^2 (\omega, n) $, $ |\Sigma'|^2 (\omega, n) $ is the spectral power of
turbulence. The presented picture of accretion is consistent with the observed variations of
X-ray and optical radiation from objects whose activity is associated with accretion of gas onto them.
\end{abstract}


\section{Introduction}
As is well known, the problem of disk accretion is that for the Keplerian rotation,
$ v_\phi \propto r^{-1/2} $, the specific angular momentum $ rv_\phi $ increases with the distance
to the center. In order for matter to fall onto the center,
angular momentum dissipation required. Taking into account the gas viscosity 
gives the necessary dissipation. Let us consider the stationary accretion of a viscous gas. Then
the $ \phi $ components of the Navier-Stokes equation yields the relation
\begin{equation}\label{NS}
\Sigma v_r \frac{1}{r}\frac {\partial} {\partial r} (rv_\phi) = \eta_s \frac{\partial} {\partial r}
\frac{1}{r} \frac{\partial}{\partial r} (rv_\phi).
\end {equation}
Here $ \Sigma = \int\rho(r, z) dz $ is the surface gas density, and
$\eta_s$ is the surface viscosity, $\eta_s = \int\eta(r, z) dz $.
Substituting from the continuity equation the relation $\Sigma v_r = -{\dot M}/2\pi r $,
we obtain the solution of (\ref{NS}) for $\eta_s = const(r)$ in the form of power-law functions,
$ v_\phi\propto r^\sigma, \, \sigma_1 = -1, \, \sigma_2 = 1 - {\dot M}/2 \pi \eta_s $.
For a disk closed to the Keplerian one, $ v_\phi \simeq r^{- 1/2}, \, \sigma_2 = -1 / 2 $, we find the value of the
accretion rate
\begin{equation}\label{Mdot}
{\dot M}\simeq 3\pi\eta_s. 
\end{equation}
From this relation it follows that
the rate of accretion is proportional to the gas viscosity $ \eta $. Without gas viscosity there is no accretion.
If we use the value of the classical viscosity of an ionized gas (plasma)
$$
\eta=0.1\left(\frac{T}{10 \, eV}\right)^{5/2}\left(\frac{10}{\Lambda_c}\right) g/s \,cm
$$
($ T $ is the plasma temperature and $ \Lambda_c $ is the Coulomb logarithm), then it turns out that
classical viscosity is not able to provide the necessary rate of accretion of
the typical value of $ {\dot M} \simeq 10^{-10} M_\odot / y \simeq 10^{16} g / s $ for observed galactic
sources, whose activity is associated with gas accretion.
Therefore, in \cite{shakura}, \cite{shakurasun} the disk model was proposed in which an anomalous
viscosity was introduced. It was assumed that the component of the tensor of viscous tensions $ \pi_{r \phi} $, which is
responsible for the radial transfer of the angular momentum, is proportional to the gas pressure $ p $,
$ \pi_{r \phi} = -\alpha p $. Models of accretion disks, based on this
assumption, are called $ \alpha $-models, and disks are called $\alpha$-discs.
Since $\pi_{r \phi} = -\eta(\partial (rv_\phi) / \partial r) / r $, then for the Keplerian disk
the introduction of the coefficient $ \alpha $ is equivalent to introduction of the anomalous viscosity $ \eta = 2 \alpha pr / v_\phi $.
Putting the thermal velocity of the gas, $ p = \rho v_T^2/2 $, and the thickness of the disk, $ h = rv_T / v_\phi $, we obtain
$ \eta = \alpha \rho v_Th $. This corresponds to the kinematic viscosity $ \nu = \alpha v_Th $.
Interpretation of this expression is as follows: there is the anomalous viscosity caused by
the turbulence of the gas flow. Then $\nu = v_t l_t / 3 $, where quantities $ v_t $ and $ l_t $ are the characteristic
velocity and scale of the turbulence respectively. Assuming $ l_t = h $ and $ v_t = \alpha v_T / 3 $, we obtain the required
expression for the turbulent viscosity.

Thus, the classical (collisional) viscosity can not provide large
accretion flows, it is necessary to introduce a turbulence. For a gas disc, in which one can neglect
the influence of the magnetic field, turbulence is an acoustic turbulence, i.e. superposition of acoustic waves
with random phases. In addition to the collisional dissipation, whose effect on the dynamics of the gas disk, as we have seen, is small,
there exists a collisionless dissipation mechanism, the well-known example of which is
the Landau damping (\cite{landau}). The Landau damping is due to the resonance interaction. Acoustic waves
in the disc also experience resonance with the azimuthal rotation. Excitation (or absorption) of waves in resonances leads to
their growth (or attenuation), i.e. to  appearance of the collisionless dissipation. This is the subject of this work. In
the second section we find resonances, in the third and fourth sections we calculate the behavior of acoustic waves in resonances,
then calculate the rate of the accretion of a gas due to the resonant interaction.

\section{Resonances}

Let us consider the motion of a gas in a thin accretion disk. Therefore, it is convenient to introduce a surface density of the matter $ \Sigma(t, r, \phi) $ through the usual density $ \rho (t, r, \phi, z) $ by the relation
$ \Sigma = \int \rho (t, r, \phi, z) dz $. Here $z$ is the coordinate orthogonal to the plane of the disk, and coordinates $ r $ and $ \phi $ are the cylindrical coordinates in the plane of the disk. In steady state
the surface density $ \Sigma $ does not depend on time and also on the azimuthal angle $ \phi $, $ \Sigma =\Sigma_0 (r) $. In an arbitrary perturbed state, the surface density is the sum of 
$\Sigma_0 $ and disturbances $ \Sigma '(t, \phi, r) $, $ \Sigma = \Sigma_0 + \Sigma' $. We will consider disturbances are not very
large, $ \Sigma '<\Sigma_0 $. In turn, the velocity of the matter in a disk has two components,
$ v_\phi (t, r, \phi) $ and $ v_r (t, r, \phi) $. The stationary velocity $ v_{\phi 0} $ is the Keplerian rotation velocity, $ v_{\phi 0} (r) = (GM / r)^{1/2} $, where $ G $ is the gravitational constant, $ M $ is the mass of the central object. Velocities also have perturbations, $ v_\phi =
v_{\phi 0} (r) + v_\phi (t, r, \phi) $, $v_r = v_r (t, r, \phi) $. We also introduce the Keplerian
frequency of rotation, $ \Omega (r) = (GM / r^3)^{1/2} $. As well as the surface density, we introduce the surface pressure of the gas, $ P = \int p(t, r, \phi, z) dz $. The pressure and the gas density in the disk is connected by the equation of state, $ p = p (\rho) $. Introducing the speed of sound, $ c_s^2 = \partial p / \partial \rho|_{\rho = \rho_0} $, the surface pressure can be represented as $ P = P_0 (r) + {\bar c}_s^2 (r) \Sigma'(t, r, \phi) $. The value of $ {\bar c}_s^2 (r) $ is the value of the square of the sound velocity at a certain middle point ${\bar z}$. 
The perturbed quantities $ v_r (t, r, \phi), \,
v_\phi (t, r, \phi), \, \Sigma '(t, r, \phi) $ can be represented as expansions
$$
(v_r, v_\phi, \Sigma')=\sum_{n=-\infty}^{\infty}\frac{1}{2\pi}\int (v_r(r,\omega,n), 
$$
$$
v_\phi(r,\omega,n), \Sigma'(r,\omega,n))\exp\{-i\omega t+in\phi\}d\omega.
$$

Equations of the ideal hydrodynamics for the two-dimensional
velocity $ v_r (r, \phi), \, v_\phi (r, \phi) $ have the form
\begin{eqnarray}\label{HD}
\frac{\partial v_r}{\partial t}+v_r\frac{\partial v_r}{\partial r}+\frac{v_\phi}
{r}\frac{\partial v_r}{\partial\phi}-\frac{v_\phi^2}{r}=-\frac{GM}{r^2}-\frac{1}
{\Sigma}\frac{\partial P}{\partial r}, \nonumber \\
\frac{\partial v_\phi}{\partial t}+v_r\frac{\partial v_\phi}{\partial r}+\frac{v_\phi}
{r}\frac{\partial v_\phi}{\partial\phi}+\frac{v_r v_\phi}{r}=-\frac{1}
{r\Sigma}\frac{\partial P}{\partial\phi}, \\
\frac{\partial\Sigma}{\partial t}+\frac{1}{r}\frac{\partial}{\partial r}
\left(\Sigma rv_r\right)+\frac{1}{r}
\frac{\partial}{\partial\phi}\left(\Sigma v_\phi\right)=0. \nonumber
\end{eqnarray}

Substituting quantities $\Sigma=\Sigma_0+\Sigma', \, v_\phi=v_{\phi 0}+v_\phi, \, v_r$ into the first two equations of the system (\ref{HD}) and linearizing, we get
\begin{eqnarray}\label{v}
v_r(r, \omega, n)=-i\frac{{\bar c_s^2}}{\Sigma_0}\frac{(\omega-n\Omega)\partial
\Sigma'/\partial r-2n\Omega\Sigma'/r}{[\omega-(n-1)\Omega]
[\omega-(n+1)\Omega]}, \nonumber \\
v_\phi(r, \omega, n)=-\frac{{\bar c_s^2}}{\Sigma_0}\frac{\Omega
(\partial\Sigma'/\partial r)/2-n(\omega-n\Omega)\Sigma'/r}
{[\omega-(n-1)\Omega][\omega-(n+1)\Omega]}.
\end{eqnarray}
In deriving equations (\ref{v}), we neglected the derivative $ (\partial \ln {\bar c}_s^2 /
\partial r) $ in comparison with the derivative $ (\partial \ln \Sigma'/
\partial r) $ since the first quantity is of the order of $ r^{-1} $, while
the second is of the order of the inverse wavelength of acoustic waves $ \lambda $, $ \lambda \simeq c_s / \omega << r $. We see that the gas velocity
strongly increases near resonances, called Lindblad resonances. And for fixed values of
$ \omega $ and $ n $ we have two resonant surfaces, $ \Omega_{-1} = \omega / (n-1) $ and
$ \Omega_{+ 1} = \omega / (n + 1) $. However,  velocities $ v_r, \, v_\phi $ are not
turn to infinity at the resonance, since poles in expressions (\ref{v}) are not at
real values of $ r $, but at complex values, since to the frequency
$ \omega $ it is necessary to add a small positive imaginary value,
$ \omega \rightarrow \omega + i0 $. This is due to the Landau's rule of bypass of the pole,
caused by the causality principle. It should be noted that resonances have the form
$ \omega = (n\pm 1) \Omega $ for the Keplerian rotation, $ \Omega \propto r^{-3/2} $.
For an arbitrary dependence $ \Omega (r) $, resonances have the following general form $ \omega =
n  \Omega \pm \kappa $, where $ \kappa $ is, so-called, epicyclic frequency
$ \kappa^2 = (2 \Omega d(r^2 \Omega) / dr) / r $. Under the Keplerian rotation $ \kappa = \Omega $.

\section{Acoustic waves}

Substituting the expressions for the perturbed velocities (\ref{v}) into the continuity equation, we obtain the equation describing density waves
\begin{eqnarray*}
&&\frac{\partial}{\partial r}\left[r{\bar c_s^2}\frac{(\omega-n\Omega)\partial
\Sigma'/\partial r-2n\Omega\Sigma'/r}{[\omega-(n-1)\Omega]
[\omega-(n+1)\Omega]}\right]+ \\  \nonumber
&&n{\bar c}_s^2\frac{\Omega
(\partial\Sigma'/\partial r)/2-n(\omega-n\Omega)\Sigma'/r}
{[\omega-(n-1)\Omega][\omega-(n+1)\Omega]}+r(\omega-n\Omega)\Sigma'=0.
\end{eqnarray*}
As before, neglecting the derivative $ \partial \ln ({\bar c}_s^2) / \partial r $ in comparison with the derivative $ \partial \ln \Sigma'/ \partial r $, but leaving the derivative $ \partial \Omega / \partial r $, which is important for resonances, we get
\begin{eqnarray}\label{eq}
&&\frac{\partial^2\Sigma'}{\partial r^2}-3\frac{\Omega}{r}\frac{n\omega-(n^2-1)\Omega}{[\omega-(n-1)\Omega][\omega-(n+1)\Omega]}
\frac{\partial\Sigma'}{\partial r}+ \\ \nonumber
&&\left[-\frac{n^2}{r^2}+\frac{3n\Omega}{r^2(\omega-n\Omega)}\frac{\omega^2-(n^2-1)\Omega^2}{[\omega-(n-1)\Omega][\omega-(n+1)\Omega]}+ \right. \\ \nonumber
&&\left. \frac{[\omega-(n-1)\Omega][\omega-(n+1)\Omega]}{{\bar c_s^2}}\right]\Sigma'=0.
\end{eqnarray}
Outside resonances, $ | \omega-n \Omega | >> \Omega $, the equation (\ref{eq}) gives the usual dispersion equation for acoustic oscillations. In the quasiclassical approximation, $ k_r r >> 1 $, $ \Sigma'\propto \exp \{i \int k_rdr \} $, we have
$$
k^2=k_r^2(r)+k_\phi^2=k_r^2(r)+\frac{n^2}{r^2}=
$$
$$
\frac{[\omega-(n-1)\Omega][\omega-(n+1)\Omega]}{{\bar c_s^2}}.
$$
We consider not global oscillations of the disk, but small-scale turbulence, that is $ k_\phi r >> 1 $, $ n >> 1 $.
Then far from resonances $ \omega = (n\pm 1)\Omega $ the right-hand side can be represented in the form $ (\omega-n\Omega)^2/{\bar c_s^2} $. Wherein
the dispersion equation has the usual form of acoustic waves propagating in a medium rotating with the frequency $ \Omega $.
It is interesting to note that if a wave propagates from internal central regions
to its resonance ($ \omega <n\Omega, \, \omega = n\Omega-k{\bar c}_s $), then its group velocity,
$ \partial \omega (r, {\bf k}) / \partial k_r = - {\bar c}_s k_r / k $, is
antiparallel to the phase velocity. While passing through the resonance region, $ \omega> n \Omega, \, \omega = n \Omega + k {\bar c}_s $, the group velocity
becomes parallel to the phase velocity, $ \partial \omega (r, {\bf k}) / \partial k_r = {\bar c}_s k_r / k $. It means that
the wave propagating from the center to the periphery before resonance has the negative radial wave vector, $ k_r <0 $. Then, having passed the
resonance, the wave vector $ k_r $ becomes positive. Approaching resonances $ \omega \simeq (n\pm 1)\Omega $, the radial wave vector $ k_r $ passes through zero. Taking into account the azimuthal
wave vector $ k_\phi = n / r $, resonances $ (k_r = 0) $ are slightly shifted, their positions are determined by the relations
$$
\omega=\Omega\left[n\pm\left(1+\frac{{\bar c_s}^2}{v_{\phi 0}^2}\right)^{1/2}\right].
$$
Since for a thin disk, $ {\bar c}_s << v_{\phi 0} $, this displacement can be neglected, and therefore the term $ n^2 / r^2 $ in Eq. (\ref{eq}) can be omitted.

Such a 'strange' behavior of the acoustic wave in the inner region ($ \omega <n \Omega $) is due to the fact that the wave energy density 
$ {\cal E} = \Sigma v^2/2 $ is not everywhere positive. In a rotating disk, the value of $ {\cal E} $ is equal to
\begin{equation*}
{\cal E}=\frac{1}{2}\Sigma_0(v_r^2+v_\phi^2)+\Sigma'v_{\phi 0}v_\phi.
\end{equation*}
If the first term is always positive,
\begin{eqnarray*}
&&\frac{1}{2}\Sigma_0(v_r^2+v_\phi^2)=\frac{{\bar c_s}^2\Sigma_0}{4}|\frac{\Sigma'}{\Sigma_0}|^2+
\frac{{\bar c_s}^4\Sigma_0}{4}\frac{n^2}{r^2(\omega-n\Omega)^2}|\frac{\Sigma'}{\Sigma_0}|^2+ \\   \nonumber
&&{\bar c_s}^4\Sigma_0\frac{n^2\Omega^2}{r^2(\omega-n\Omega)^4}|\frac{\Sigma'}{\Sigma_0}|^2+\frac{{\bar c_s}^2\Sigma_0}{16}\frac{\Omega^2}{(\omega-n\Omega)^2}
|\frac{\Sigma'}{\Sigma_0}|^2,
\end{eqnarray*}
and under conditions $ n >> 1, \, {\bar c}_s << v_{\phi 0} $ is equal to
$$
\frac{{\bar c_s}^2\Sigma_0}{4}|\frac{\Sigma'}{\Sigma_0}|^2,
$$
then the summand 
$$
\Sigma'v_{\phi 0}v_\phi=\frac{{\bar c_s}^2\Sigma_0}{2}\frac{n\Omega}{\omega-n\Omega}|\frac{\Sigma'}{\Sigma_0}|^2
$$
is negative in the inner region $ \omega <n \Omega $. As a result, the expression for the wave energy density has the form
\begin{equation}\label{energy}
{\cal E}=\frac{{\bar c_s}^2\Sigma_0}{4}\frac{\omega+n\Omega}{\omega-n\Omega}|\frac{\Sigma'}{\Sigma_0}|^2.
\end{equation}
In deriving the expression for the wave energy density, we used formulas (\ref{v}) for the radial and azimuthal velocities
$ v_r $ and $ v_\phi $, where the denominator was replaced by $ (\omega-n \Omega)^2 $, and also used the expression $ k_r = \pm (\omega-n \Omega) / {\bar c}_s $, which is valid far from resonances.
Here one also need to keep in mind that the product  $a(\Sigma')^2$ means $(a\Sigma'\Sigma'^*+a^*\Sigma'^*\Sigma)/4$, where the sign ($^*$) denotes a complex conjugation.

Thus, we see that for acoustic waves with $ \omega <n \Omega_{max} $ their energy is negative in the inner region $ \Omega(r)> \omega / n, \, r> r_{min},
\, \Omega_{max} = \Omega(r_{min}) $. This means that such waves are easily excited in a dissipative medium by so-called dissipative instability (\cite{mich}).
This is because any dissipation leading to decrease of the total energy means growth of the amplitude of the wave of negative energy.

\section{Inverse Landau damping}

Famous Landau damping, which was discovered by him in equilibrium
plasma (\cite{landau}), means a collisionless phenomenon due to the resonant interaction of the wave with particles. It leads to wave damping. However, in a non-equilibrium medium, for example, a beam of fast particles in plasma, certain waves will, on the contrary, grow. Decrement is replaced by increment. Thus, the term 'inverse'  means not decrease, but increase of the wave amplitude, and this phenomenon has the same collisionless character like damping.

Near resonances $ \omega = (n \pm 1) \Omega $ the quasiclassical radial wave vector $ k_r $, vanishing at resonance points, becomes purely imaginary inside the interval between them, $ k_r^2 <0 $. This occurs in the range $ r_{-1} <r <r_{+1} $, where values of $ r_{\pm 1} $ are determined the relations
$ \omega = (n \pm 1) \Omega (r_{\pm 1}) $. For $ n >> 1 $, points $ r_{\pm 1} $ are located close to the point
$ r = r_0, \, \omega= n\Omega (r_0) $, $ r_{\pm 1} = r_0 (1 \pm 2 / 3n) $. As a result, we have
\begin{equation*}
k_r^2=-\frac{9}{4}\frac{\omega^2}{{\bar c}_s^2}\frac{(r-r_{-1})(r_{+1}-r)}{r_0^2}.
\end{equation*}
The purely imaginary value of the radial wave vector means amplification or attenuation of waves when they pass through resonance region 
$ r_{-1} <r <r_{+1} $. The gain (or attenuation) $ A $ is equal to
\begin{eqnarray}\label{A}
&&|A|=\exp{\Lambda}, \, \Lambda=\pm\int_{r_{-1}}^{r_{+1}}|k_r|dr= \\  \nonumber
&&\pm\frac{3\pi\omega}{16{\bar c}_s r_0}(r_{+1}-r_{-1})^2=
\pm\frac{\pi}{3}\frac{\omega r_0}{{\bar c}_s n^2}.
\end{eqnarray}
For not too large $ n $, $ n <(\omega r_0 / {\bar c}_s)^{1/2} $, the wave greatly changes its amplitude, passing through the resonance
region. For large values of $ n $, the distance between the points $ r_{\pm 1} $ becomes less than the characteristic wavelength in the resonance region,
and the resonant layer becomes transparent for the wave. We are interested in the case of opacity. In order to determine 
whether amplification or attenuation of waves happens, it is necessary to solve the equation (\ref{eq}) in the resonance region $ r_{-1} <r <r_{+1} $. The problem is posed as follows: a wave with the negative radial wave vector, $ k_r <0 $, but with the positive group velocity, $ d \omega / dk_r> 0, \, k_r \simeq (\omega-n \Omega) / {\bar c}_s $, is incident onto the 
layer $ r_{-1} <r <r_{+1} $ from the left side, i.e. from inner regions of the disk. Part of a wave can be reflected from the layer,
part passes through the layer and begins to propagate into the outer region $ r> r_{+1} $, $ k_r> 0 $. The schema is shown in Figure 1.
Thus, in the region $ r> r_{+ 1} $ there exist only the past wave exists, whereas in the region $ r <r_{-1} $ there are both the incident and the reflected wave.
We start, naturally, from the region $ r> r_{+ 1} $. In the neighborhood of $ r \simeq r_{+1} $ the equation (\ref{eq}) takes the form

\begin{figure}
\centering
\includegraphics[width=6cm]{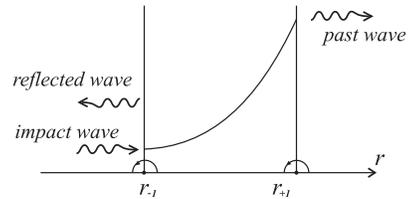}
\caption{The scheme of the passage of the acoustic wave, incident from the left, through the resonant layer. There are also the reflected and the past waves. The past wave exponentially increases. Semicircles above the poles show the rule of their bypass.}
\end{figure}

\begin{equation}\label{eq1}
\frac{\partial^2 \Sigma'}{\partial r^2}-\frac{1}{r-r_{+1}}\left(\frac{\partial \Sigma'}{\partial r}-\frac{2n}{r_0}\Sigma'\right)+
\frac{3\omega^2}{{\bar c}_s^2n}\frac{r-r_{+1}}{r_0}\Sigma'=0.
\end{equation}
From this equation we see that the characteristic size $\lambda$ of the change of the wave amplitude $ \Sigma'$ is 
$\lambda \simeq r_0 ({\bar c}_s^2n/3 \omega^2r_0^2)^{1/3}$. Then, under parameters $ n <(\omega r_0 / {\bar c}_s)^{1/2} $ we can neglect the 
term $ (2n / r_0) \Sigma'$ in this equation. Finally we obtain
$$
\frac{\partial^2 \Sigma'}{\partial r^2}-\frac{1}{r-r_{+1}}\frac{\partial \Sigma'}{\partial r}+
\frac{3\omega^2}{{\bar c}_s^2n}\frac{r-r_{+1}}{r_0}\Sigma'=0.
$$
Introducing the dimensionless coordinate 

$x =(3\omega^2r_0^2/{\bar c}_s^2 n)^{1/3}(r-r_{+ 1})/r_0 $, 
we obtain the equation
\begin{equation}\label{eqx}
\frac{\partial^2\Sigma'}{\partial x^2}-\frac{1}{x}\frac{\partial\Sigma'}{\partial x}+x\Sigma'=0.
\end{equation}
The required solution of the equation (\ref{eqx}) for $ x> 0 $ is
\begin{equation}\label{H1}
\Sigma'=a xH^{(1)}_{2/3}\left(\frac{2}{3}x^{3/2}\right).
\end{equation}
Here the function $ H^{(1)} $ is the Hankel function of the first kind. It describes a wave traveling in the positive radial direction.
Its amplitude at $ x = 0 $ is equal to $ -ia3^{2/3} \Gamma(2/3) / \pi $. At large distances from the resonance, $ x> 1 $, the solution goes to the
quasiclassical wave with the radial wave vector $ k_r \propto x^{1/2} $, $ \Sigma '= a (3 / \pi)^{1/2} x^{1/4} \exp \{i ( 2x^{3/2} /3-7\pi / 12) \} $.
Then, at the distance $ r-r_0 \simeq r_0 / n << r_0 $, when the influence of resonances $  \omega = (n \pm 1) \Omega $ weakens,
the acoustic wave becomes the wave with the 'ordinary' dispersion, $ k_r = (\omega-n \Omega) / {\bar c}_s $.

Now we need to analytically continue the solution (\ref{H1}) into the region $ r <r_{+1} $, $ x <0 $. Recall that the pole $ \omega + i0- (n + 1) \Omega (r) = 0 $ is not on the real axis $ r $, but under it in the complex plane $ (Re(r), Im (r)) $, since $ d \Omega / dr <0 $. Thus, the bypass of the 
point $ r = r_{+ 1} $
must be realized counterclockwise in the upper half-plane $ Im (r)> 0 $ (analogous to the Landau's rule of bypass of a pole, but in the lower half-plane).
Thus, $ x = | x | \exp (-i \pi) $, where $ | x | \propto r_{+1} -r> 0 $. The Hankel function of the first kind $ H^{(1)}_{2/3} $ goes to the McDonald 
function $ K_{2/3} $,
$ H^{(1)}_{2/3} (2x^{3/2} / 3, x <0) = - (2i / \pi) \exp (-i \pi / 3) K_{2 / 3} (2 | x |^{3/2} / 3) $ (\cite{handbook}). Therefore, the solution in the region 
$ r_{-l} <r <r_{+ 1} $ is
\begin{equation}\label{K}
\Sigma'=\frac{2ia}{\pi}\exp(-i\pi/3)|x|K_{2/3}\left(\frac{2}{3}|x|^{3/2}\right).
\end{equation}
When passing through the boundary $ r = r_{+ 1} $, the amplitude of the wave does not change, but acquires the additional phase $ 2 \pi / 3 $. Thus, in the region of the resonance the wave ceases to oscillate, and begins to decay exponentially when moves away from the point $ r = r_{+ 1} $ and approaches 
the point $ r = r_{-1} $. It is necessary to match the solution obtained with a solution near other resonance point $ r = r_{-l} $. By the same way as
before, we introduce the dimensionless coordinate $ y = (3 \omega^2r_0^2 / {\bar c}_s^2 n)^{1/3} (r-r_{-1}) / r_0 $ and transform the equation (\ref{eq}). We get
\begin{equation*}
\frac{\partial^2\Sigma'}{\partial y^2}-\frac{1}{y}\frac{\partial\Sigma'}{\partial y}-y\Sigma'=0.
\end{equation*}
Solutions of this equation are both the MacDonald function, $ yK_{2/3} (2y^{3/2} / 3) $, and the modified Bessel function, $ yI_{2/3} (2y^{3/2} / 3) $. To match them with the function $ K_{2/3} $ that exponentially falls with $ |x| $, it is necessary to choose the function $ I_{2/3} $ exponentially growing with $ y $,
\begin{equation}\label{eqy}
\Sigma'=b yI_{2/3}\left(\frac{2}{3}y^{3/2}\right).
\end{equation}
Equating the asymptotic values of $ \Sigma'$ (\ref{K},\ref{eqy}) and its derivatives with respect to the radius $ r $ at some point $ r^*, \, r_{-l} <r^* <r_{+1} $, which turns the middle of the segment $(r_{-1}, r _{+ 1}), r^* = (r_{+1} + r_{-l}) / 2 = r_0 $, we obtain the connection between amplitudes $ a $ and $ b $  
$$
b=2a\exp(i\pi/6)\exp\left(-\frac{2^{7/2}}{9}\frac{\omega r_0}{{\bar c}_sn^2}\right).
$$
Finally, we extend analytically the solution of (\ref{eqy}) into the region $ r <r_{-1}, \, y <0 $. As we have already established, it is necessary to put 
$ y = | y | \exp (-i \pi) $ in the expression (\ref{eqy}), then $ I_{2/3} (2y^{3/2} / 3, y <0) = \exp (i \pi / 3) J_{2/3} (2 | y |^{3/2} / 3) $ (Abramowitz \& Stegun, 1964). Therefore, the solution in the region $ r <r_{-1} $ is
\begin{equation}\label{H12}
\Sigma'=-\frac{b}{2}\exp(i\pi/3)|y|\left(H^{(1)}_{2/3}(\frac{2}{3}|y|^{3/2})+H^{(2)}_{2/3}(\frac{2}{3}|y|^{3/2})\right).
\end{equation}
Here the Hankel function of the first kind, $ H ^{(1)}_{2/3} $, describes a wave incident onto a resonant layer from the inner region of the disk $ r <r_{-1} $. It has a negative wave vector, $ k_r <0 $. Using the connection between quantities $ a $ and $ b $, we find that the amplitude of the past wave $ a $ exponentially increases in comparison with the amplitude of the incident wave $ -b \exp (i \pi / 3) / 2 $. Its amplification is
$$
A=\exp(i\frac{\pi}{2})\exp\left(\frac{2^{7/2}}{9}\frac{\omega r_0}{{\bar c}_sn^2}\right).
$$
We note that the exponent obtained is in good agreement with the expression for $ \Lambda $ (\ref{A}) found from the quasiclassical
approximation, $ 2^{7/2} / 9 = 1.2 \pi / 3 $. In the region $ r <r_{-1} $, there is also a reflected wave described by the Hankel function of the second
kind, $ H^{(2)}_{2/3}, \, k_r> 0, \, d \omega / dk_r <0 $. Its amplitude is equal to the amplitude of the incident wave.

Thus, we see that acoustic waves having not too large azimuthal numbers $ n $, $ n <(\omega r_0 / {\bar c}_s)^{1/2} $,
passing though resonant points, experience exponential growth. Since the value of $ r_0 $ itself depends on $ n $, $ \Omega (r_o) = \omega / n $,
then the resonance amplification condition is as follows
$$
n<n^*=\left(\frac{v_{\phi \, max}}{{\bar c}_s}\right)^{3/4}\left(\frac{\omega}{\Omega_{max}}\right)^{1/4}.
$$
Values with the index 'max' correspond to their values at the inner edge of the disk. On the other hand, for a resonance to exist, it is necessary to be
$ n> \omega / \Omega_{max} $. Thus, a high level of turbulence can be for waves with azimuthal wave numbers lying in the range
\begin{equation}\label{n*}
\frac{\omega}{\Omega_{max}}<n<\left(\frac{v_{\phi \, max}}{{\bar c}_s}\right)^{3/4}\left(\frac{\omega}{\Omega_{max}}\right)^{1/4},
\end{equation}
which is possible for not too large frequencies, $ \omega <\omega^* = \Omega_{max} (v_{\phi \, max} / {\bar c}_s) $. Waves of this frequency 
band, $ \omega <\omega^* $,
and from the azimuthal wave number region, $ n <n^* $, should initially have huge amplitudes, $ \Sigma'\simeq \Sigma_0 $, since for them the coefficient
of the amplification, $ | A | = \exp \Lambda, \, \Lambda> 1 $, is exponentially large. It is clear that their nonlinear interaction, decays and fusions of different harmonics, as well as the inverse
influence onto the rotation profile of the disk $ \Omega (r) $,
will lead to the formation of a wide range of turbulence with the most probable power-law distribution over frequencies $ \omega $ and wave numbers $ n $,
$ | \Sigma'/ \Sigma_0 |^2 (n, \omega) \propto \omega^{- \beta} \, n^{- \gamma} $.

It should be noted that, just as in a plasma with the inverse Landau damping (for example, beam instability), the energy of the waves is drawn from the energy
of motion of the matter. In our case it is from the rotation of the disk. Under the Keplerian rotation, $  d \Omega (r) / dr <0 $, the amplification of acoustic waves, propagating
out, should lead to slow down of the rotation of the inner layers of the disk, i.e. equalizing of the angular velocity of rotation. So, for solid rotation,
$ \Omega (r) = const $, the amplification effect is absent, and for $ d \Omega (r) / dr> 0 $ the inverse Landau damping is replaced by the Landau damping, which corresponds to
the displacement of the pole in the complex region $ (Re (r), Im (r)) $ from the upper half-plane to the lower half-plane. Slowing down of the rotation of the disk, associated with the excitation of
acoustic waves, leads to the decrease of the angular momentum of the disk and to possibility of an accretion.

\section{Accretion}

In an acoustic wave propagating from the inner edge of the disk to the periphery, $ d \omega / dk_r> 0, \, k_r = (\omega-n \Omega) / {\bar c}_s $, the matter moves in the radial direction with the velocity proportional to the wave amplitude $ \Sigma'$. Far from resonances, the radial velocity 
is (see the equation (\ref{v})),
$$
v_r=-i\frac{{\bar c}_s^2}{\Sigma_0}\left[\frac{\partial\Sigma'/\partial r}{(\omega-n\Omega)}-\frac{2n\Omega\Sigma'/r}{(\omega-n\Omega)^2}\right].
$$
The mass intersecting a circle of radius $ r $ per unit time equals to
$$
{\dot M}=r\int_0^{2\pi}d\phi v_r(r,\phi)\Sigma(r,\phi)=r\int_0^{2\pi}d\phi(v_r\Sigma'+v_r^{(2)}\Sigma_0)
$$
$$
={\dot M_1} +{\dot M_2}.
$$
Here, the velocity $ v_r $ is the radial velocity of the matter (\ref{v}), proportional to the first power of the amplitude $ \Sigma'$. The radial velocity 
$ v_r^{(2)} $ is the second-order velocity, proportional to the square of the amplitude $ \Sigma'^2 $. We first calculate the value of $ {\dot M}_1 $.
$$
{\dot M}_1(\omega)=r\frac{1}{2\pi}\int_0^{2\pi}d\phi
$$
$$
\int d\omega' \sum_{n,n'}v_r(\omega',n)\Sigma'(\omega-\omega',n')
$$
$$
e^{i(n+n')\phi}=r\int d\omega'\sum_n v_r(\omega',n)
\Sigma'(\omega-\omega',-n).
$$
Since $ \Sigma'(- \omega, -n) = \Sigma'^* (\omega, n) $, we obtain
$$
{\dot M}_1(\omega)=r\int d\omega'\sum_n v_r(\omega',n)
\Sigma'^*(\omega'-\omega,n).
$$
Here the sign '$*$' means complex conjugation. The radial gradient $ \partial \Sigma'/ \partial r $ is equal to 
$ ik_r \Sigma'= i (\omega-n \Omega) \Sigma' / {\bar c} _s $. Recall that in the inner region, $ \omega <n \Omega $, the radial wave vector is negative for a wave propagating in a positive direction, $ d \omega / dk_r> 0 $. As a result, we have
$$
{\dot M}_1(\omega)=\frac{r{\bar c}_s}{\Sigma_0}\int d\omega'\sum_n 
\left[1+\frac{2in\Omega{\bar c}_s}{r(\omega'-n\Omega)^2}\right]
$$
$$
\Sigma'(\omega',n)\Sigma'^*(\omega'-\omega,n).
$$
We will now assume that the acoustic waves are a random turbulent field, i.e. quantities $ \Sigma'$ contain random phases. In this case 
averaging $ \langle ... \rangle $ over realization of a random field gives
$$
\langle\Sigma'(\omega')\Sigma'^*(\omega'-\omega)\rangle=2\pi|\Sigma'|^2(\omega')\delta(\omega).
$$
The quantity $ |\Sigma'|^2 (\omega') $ is the spectral density of turbulence. Since the spectral density is an even function of  
arguments, $ | \Sigma |^2 (- \omega, -n) = | \Sigma |^2 (\omega, n) $, then we can restrict ourselves only to positive frequencies, $ \omega> 0 $, and to positive azimuthal wave numbers, $ n> 0 $. Thus, the value of $ {\dot M}_1 = \int {\dot M}_1 (\omega) d \omega $, that is the part of the accretion rate, is positive for waves propagating outward from internal areas of the disk, and is equal to
\begin{equation*}
{\dot M}_1=2\pi r{\bar c}_s\Sigma_0\int_0^\infty\sum_{n\geq 0}\Big{|}
\frac{\Sigma'}{\Sigma_0}\Big{|}^2(\omega',n)d\omega'.
\end{equation*}
In order to calculate the second part of the accretion rate $ {\dot M}_2 = r \int_0^{2 \pi} d \phi v_r^{(2)} \Sigma_0 $ it is necessary to find the radial velocity of the second order $ v_r^{(2)} $. Equations for second-order quantities $ v_r^{(2)}, \, v_\phi^{(2)}, \, \Sigma^{(2)} $ follow from the system of equations (\ref{HD}):
\begin{eqnarray}\label{second}
&&\frac{\partial v_r^{(2)}}{\partial t}+\Omega\frac{\partial v_r^{(2)}}{\partial\phi}-2\Omega v_\phi^{(2)}+
\frac{{\bar c}_s^2}{\Sigma_0}\frac{\partial\Sigma^{(2)}}{\partial r}=   \nonumber \\
&&-v_r\frac{\partial v_r}{\partial r}-
\frac{v_\phi}{r}\frac{\partial v_r}{\partial\phi}+\frac{v_\phi^2}{r} \nonumber \\
&&-\frac{1}{\Sigma_0}\left(\int\rho_0dz
\int\rho'\frac{\partial\rho'}{\partial r}\frac{\partial c_s^2}{\partial\rho_0}dz-
\int\rho'dz\int c_s^2\frac{\partial\rho'}{\partial r}dz\right); \nonumber \\
&&\frac{\partial v_\phi^{(2)}}{\partial t}+\Omega\frac{\partial v_\phi^{(2)}}{\partial\phi}+
\frac{1}{r}\frac{\partial(r^2\Omega)}{\partial r}v_r^{(2)}+\frac{{\bar c}_s^2}{r\Sigma_0}\frac{\partial\Sigma^{(2)}}{\partial\phi} = \nonumber \\
&&-\frac{v_r}{r}\frac{\partial (rv_\phi)}{\partial r}-
\frac{v_\phi}{r}\frac{\partial v_\phi}{\partial\phi} -\nonumber \\
&&\frac{1}{\Sigma_0}\left(\int\rho_0dz
\int\rho'\frac{1}{r}\frac{\partial\rho'}{\partial\phi}\frac{\partial c_s^2}{\partial\rho_0}dz-
\int\rho'dz\int c_s^2\frac{1}{r}\frac{\partial\rho'}{\partial\phi}dz\right);  \nonumber \\
&&\frac{\partial\Sigma^{(2)}}{\partial t}+\Omega\frac{\partial\Sigma^{(2)}}{\partial\phi}+
\frac{1}{r}\frac{\partial}{\partial r}\left(r\Sigma_0 v_r^{(2)}\right)+
\frac{\Sigma_0}{r}\frac{\partial v_\phi^{(2)}}{\partial\phi}= \\
&&-\frac{1}{r}\frac{\partial}{\partial r}\left(r\Sigma' v_r \right)-\frac{1}{r}\frac{\partial}{\partial\phi}\left(\Sigma' v_\phi\right). \nonumber
\end{eqnarray}
Here we are interested in the radial velocity $ {\bar v}_r^{(2)} $, which does not depend on the time $ t $ and on the azimuth angle $ \phi $, and is the only one which gives contribution to the second part of the accretion rate $ M_2 $. From the second equation of system (\ref{second}) it follows that
\begin{equation}\label{v2}
{\bar v_r^{(2)}}= -v_r\frac{\partial (rv_\phi)}{\partial r}/\frac{\partial(r^2\Omega)}{\partial r}.
\end{equation}
We see that the mean radial velocity of the second order $ {\bar v}_r^{(2)} $ appears as compensation of acceleration of the matter in the azimuthal direction caused by the quadratic action of velocities $ v_r, \, v_\phi $ of the acoustic wave. Substituting values of 
velocities $ v_r, \, v_\phi $ from (\ref{v}) into the expression (\ref{v2}), we obtain
$$
{\dot M}_2=-{\dot M}_1-\pi{\bar c}_s r\Sigma_0\left(\frac{{\bar c}_s}{v_{\phi 0}}\right)^2\int_0^\infty\sum_{n\geq 0}\frac{n^2\Omega^2}{(\omega-n\Omega)^2}
$$
$$
\Big{|}\frac{\Sigma'}{\Sigma_0}\Big{|}^2(\omega,n)d\omega
$$
The expression obtained agrees with the relation $ \Sigma_0 {\bar v}_r^{(2)} = - \Sigma'v_r + const (r) / 2 \pi r $, which follows from third equation of the system (\ref{second}). Therefore, $ {\dot M}_2 = 2 \pi r \Sigma_0 v_r^{(2)} = - {\dot M}_1 + const (r) $. Since in our calculations we have neglected  derivatives of slowly varying quantities $ {\bar c}_s, \, \Sigma_0 $ with respect to the radius $ r $, then they can be considered as constants. Finally we have
\begin{equation}\label{Mdot2}
{\dot M}=-\pi{\bar c}_s r\Sigma_0\frac{{\bar c}_s^2}{v_{\phi 0}^2}\int_0^\infty\sum_{n\geq 0}\frac{n^2\Omega^2}{(\omega-n\Omega)^2}
\Big{|}\frac{\Sigma'}{\Sigma_0}\Big{|}^2(\omega,n)d\omega.
\end{equation}
Here the sign minus means that acoustic waves propagating outward from the inner edge of the disk, where $ k_r = (\omega-n \Omega) / {\bar c}_s $, 
induce the opposite motion of the matter of the disk, i.e. its accretion. This is due to the fact that the wave with fixed values  of $\omega$ and $n$,  propagating in the positive direction $ r $, increases its energy (\ref{energy}), since the Keplerian rotation velocity $ \Omega $ decreases with 
increasing of $ r $. This is true
as for the inner region, where $ \omega <n \Omega $ and the wave has the negative energy, and in the external one, $ \omega> n \Omega $, where the energy is positive.
Thus, carrying out of acoustic waves outside
should be accompanied by decrease of the energy of the matter of the disc. Since the energy per unit mass of the matter is negative and is 
equal to $ -v_{\phi 0}^2/2 \propto r^{-1} $
for the Keplerian rotation, then decrease of the energy means motion toward the center. It should be noted that acoustic waves propagating from the external
edge of the disk to the center, $ k_r = - (\omega-n \Omega) / {\bar c} _s $, cause the outflow of the matter, $ {\dot M}> 0 $. The expression for $ {\dot M} $ in this case has the same form as (\ref{Mdot2}) but with the sign plus.

Let us determine the quantity entering into expression (\ref{Mdot2}) as the effective dimensionless power of acoustic turbulence $ W $,
$$
W=\int_0^\infty\sum_{n\geq 0}\frac{n^2\Omega^2}{(\omega-n\Omega)^2}\Big{|}
\frac{\Sigma'}{\Sigma_0}\Big{|}^2(\omega,n)d\omega.
$$
Then the expression for the accretion rate (\ref{Mdot2}) becomes $ |{\dot M}| =\pi r {\bar c}_s \Sigma_0 ({\bar c}_s / v_{\ phi 0})^2 W $. Therefore, comparing the formula (\ref{Mdot2}) with the expression (\ref{Mdot}) and introducing the thickness of the disk, $ h = r {\bar c}_s / v_{\phi 0} $, one can define the turbulent kinematic viscosity, $ \nu = h (h / r) {\bar c}_s W / 3, \, | {\dot M} | = 3\pi \nu\Sigma_0 $. Thus, the characteristic scale and the turbulent velocity are quantities $ l_t = h, \, v_t = {\bar c}_s (h / r) W$ respectively, $\nu = l_t v_t / 3 $. This determines the value of $ \alpha $,
\begin{equation}\label{alpha}
\alpha=\frac{h}{r}W=\frac{h}{r}\int_0^\infty\sum_{n\geq 0}\frac{n^2\Omega^2}{(\omega-n\Omega)^2}\Big{|}
\frac{\Sigma'}{\Sigma_0}\Big{|}^2(\omega,n)d\omega.
\end{equation}
We see that in the case of acoustic turbulence, the parameter $ \alpha $ is uniquely determined by the level of turbulence $ W $.
It should be noted that the presence of the denominator $ (\omega-n \Omega)^2 $ in the expression (\ref{alpha}) for $ W $ does not mean that the resonance gives
infinite contribution to the integral (\ref{alpha}).  Expressions obtained are in the quasiclassical approximation, which does not work in
the region $ k_r \simeq 0 $. We used the expression $ k_r = \pm (\omega-n \Omega) / {\bar c}_s $. In fact, the resonance occurs in
region between points $ r_{- 1}, \, \omega = (n-1)\Omega (r_{- 1}) $, and $ r_{+ 1}, \, \omega = (n + 1) \Omega (r_{+ 1}) $. The distance between
points $ r_{+ 1} -r_{-1} = \Delta r = 4r / 3n << r $ determines the minimum value of the wave vector, $ k_ {min} \simeq \Delta r^{- 1} $.
This means that $ | \omega-n \Omega |> 3 ({\bar c}_s / v_{\phi 0}) n \Omega / 4 $. 

Summation over azimuthal numbers $ n $ can be
replaced by integration, and the expression for $ W $ can be reduced to the form
$$
W=\int\limits_0^{\infty}d\omega\dashint_1^{\infty}
\frac{d n}{(n-\omega/\Omega)}\frac{\partial}{\partial n}\left[n^2\Big{|}\frac{\Sigma'}{\Sigma_0}\Big{|}^2(n,\omega)\right].
$$
The integral in the right-hand side is the integral in the sense of the principal value and is the Hilbert transformation. It is a power-law function of $ \omega / \Omega $ with the same exponent as the integrand function $ \partial n^2 | \Sigma '|^2 (n) / \partial n $ if it is a power-law function of the argument $ n $. Assuming $ | \Sigma'/ \Sigma_0 |^2 (n, \omega)= W_0 (\omega) n^{- \gamma} $, we obtain
$$
W=(2-\gamma)\cot(\pi\gamma)\int_0^\infty W_0(\omega)\left(\frac{\omega}{\Omega}\right)^{1-\gamma}d\omega, \, \gamma>1.
$$
The accretion rate onto the star $ {\dot M}_a $ is determined by the value of $ | {\dot M} | $ at $ r = r_{min} $,
\begin{equation}\label{a}
{\dot M}_a=\pi r{\bar c}_s\Sigma_0 ({\bar c}_s/v_{\phi 0})^2 W\Big{|}_{r=r_{min}}.
\end{equation}
Since, as we see, $ W \propto \Omega^{\gamma-1} $, for stationary accretion it is necessary to establish such distribution of the spectrum of the acoustic turbulence $ | \Sigma '/ \Sigma_0 |^2 (r) $ over the radius $ r $, so that the product $ r {\bar c}_s \Sigma_0 ({\bar c}_s / v_{\phi 0})^2 \Omega^{\gamma-1} W_0 (r) $ remains constant. Such dependence of the turbulence power on the radius for stationary accretion indicates that, generally speaking, the 
quantity $ \alpha = {\bar c}_sW / v_{\phi 0} $ is not constant along the disk.

The picture of turbulent accretion presented here can be to approved with
observations of variations in the X-ray flux from accreting
sources. It is natural to assume that the change in the radiation flux is proportional to
the rate of the accretion of the matter $ {\dot M}_a (\omega) $.
Then we see that the observed power-law spectrum
$ {\dot M}_a (\omega) \propto \omega^{-\beta'} $ reflects the power-law
spectrum of acoustic waves $ |\Sigma'/ \Sigma_0 |^2 (\omega) \propto \omega^{-\beta}, \, \beta = \beta' + 1- \gamma $. For different sources
the value of $ \beta '$ is of the order of unity. So for SS433 $ \beta'= 1.5 $ in a wide range of
frequencies $ f $ from $ 10^{-7} \, Hz $ to $ 10^{- 1} \, Hz $ both in the X-ray and in the optical
ranges of the spectrum of electromagnetic waves (\cite{revet}).

\section{Discussion}

The study of the propagation of acoustic waves in a rotating medium has been the subject of many studies since the 1970s (\cite{gl}, \cite{gt}, \cite{drury1}, \cite{drury2}, \cite{pp1}, \cite{pp2},
\cite{pp3}, \cite{ngg}, \cite{glatzel1}, \cite{glatzel2}).
The present interpretation of the interaction of waves with a resonant layer is as follows. The $ r_{-1}<r<r_{+1} $ layer is a barrier for
acoustic waves propagating from the inner regions of the disk to the resonant region. The radial wave number becomes purely imaginary inside the layer.
The wave cannot propagate there, and the region $ r_{-1}<r<r_{+1} $ is declared forbidden for acoustic waves. They can experience there only the subbarrier tunneling onto the allowed domain $r> r_{+ 1} $. Therefore, the amplitude of the wave transmitted through the layer is exponentially small compared to the amplitude of the incident wave, and the reflected wave can slightly increase only. This conclusion is mainly based on the consideration of a model problem in which the motion of a liquid (gas) is considered in the frame rotating with the angular frequency $\Omega(r_0)$ (see, for example, papers of \cite{gt} and \cite{ngg}).  Here $\Omega(r_0)$ is the rotation frequency of the disk in the middle of the resonant layer. Further, the unperturbed rotation frequency $ \Omega $ is assumed to be linearly falling with the coordinate $ r-r_0 $, and the cylindrical coordinates $ (r, \phi) $ are replaced by Cartesian coordinates $ (r, y), \, y = r\phi $. Moreover, the equation (\ref{eq}) for acoustic waves is simplified, only the first and the last terms in left hand side of it remain, 
\begin{equation}\label{eq2}
\frac{\partial^2\Sigma'}{\partial r^2} + \frac{[\omega-(n-1)\Omega][\omega-(n+1)\Omega]}{{\bar c_s^2}}\Sigma'=0.
\end{equation} 
For a linear dependence of the rotation frequency on the coordinate $r, \, \Omega-\Omega(r_0) \propto -(r-r_0) $, the equation (\ref{eq2}) reduces to the standard equation
\begin{equation}\label{pcy}
\frac{\partial^2\Sigma'}{\partial z^2}+\left[\frac{z^2}{4} - C\right]\Sigma'=0.
\end{equation}
Here the coordinate $z$ is, $ z = (3\omega r_0 / {\bar c_s})^{1/2} (r-r_0) / r_0 $ and the constant $C$ is, $ C = \omega r_0 / 3 {\bar c_s} n^2 $ .
The equation (\ref{pcy}) is the parabolic cylinder differential equation. It has two fundamental solutions, $ U (C, z) $ and $ U (C, -z) $. 
The first, $ U (C, z) $, exponentially falls passing through the region $-2C^{1/2} <z <2C^{1/2} $, another, U (C, -z), independent of $ U (C, z) $, grows exponentially during the transition from
$ z <-2C^{1/2} $ to $ z> 2C^{1/2} $. The dependence of solutions $ U (C, z), \, U (C, -z) $ on the coordinate $ z $ are shown on Figure 2. We dwell here  onto mathematics in details since its misunderstanding leads to mistakes. Using real independent functions 
$ U (C, z), \, U (C,-z) $ one can build two independent complex functions $ E(z) $ and $ E^*(z) $ (\cite{handbook}),
$$
E(C,z)=k^{-1/2}U(C,z)+ik^{1/2}U(C,-z), 
$$
$$
k=(1+\tau^{-2})^{1/2} -\tau^{-1}, \, \tau=\exp(-\pi C);
$$
$$
E^*(C,z)=k^{-1/2}U(C,z)-ik^{1/2}U(C,-z).
$$ 
It should be noted that $ \pi C = |\Lambda| $ (7). Functions $E(z), E^*(z)$ are convenient to use because for large arguments $ z $, $ |z| >> 1 $, they become propagating quasiclassical waves, $ E(C, z) \propto \exp (iz^2/4) $, $ E^*(C, z) \propto \exp (-iz^2/4) $. Any third solution other than $ E, \, E^* $ can be represented as a superposition of these two. In particular, the handbook (\cite{handbook}) gives the relationship between waves 
$ E (z), E^* (z), E^*(-z) $,
$$ 
E^*(C,z)-(1+\tau^2)^{1/2}E(C,z)=-i\tau E^*(C,-z).
$$
For small values of $\tau$, i.e. large values of $ \pi C $, the right side in this ratio is much less than the left. Authors \cite{ngg}
on the basis of this relation made the following conclusion.
The quote (page 10): 'This equation has a particularly transparent physical interpretation. It say the ingoing wave $E^*(z)$ of unit amplitude interacts
with the forbidden region around corotation to produce a transmitted wave of amplitude $\tau$ and a reflected wave of amplitude $(1+\tau^2)^{1/2}$.'
Well, this is the solution falling with $ z $. And where is a growing solution conjugated to it? After all, the real function $ U (C, -z) $, corresponding to 
$ U (C, z) $, grows exponentially. Let's choose another triple of solutions: $ E (z), E (-z), E^*(-z) $. $E (z)$ is the wave that  passes through the resonant layer in the positive direction of $ z $, $ E(-z)$ is the wave incident on the layer on the left, and $ E ^ * (- z) $ is the wave reflected from the layer. The relation between these waves is not difficult to obtain using expressions of $ E(z), \, E^*(z) $, defined above, in terms of $ U (z), \, U (-z) $. We get
\begin{equation}\label{c2}
E(C,z)=i\tau^{-1}\left[((1+\tau^2)^{1/2}E^*(C,-z)-E(C,-z)\right].
\end{equation}
We see that the transmitted wave $E(z)$ is $ \tau^{-1} $ times larger than the incident one $E(-z)$. This is the exponentially growing solution. But it is not written in the handbook \cite{handbook}. And its interpretation
does not match the above quote. The relation (\ref{c2}) just corresponds to the solution obtained above in this paper. Formally, mathematics does not make a selection between a falling and a growing solutions, they are equivalent. How to choose  a solution corresponding to the physical problem? It is necessary to use the physical principle of causality, which says that only the past affects the present. In order for the contribution of the past not to be infinite, it is necessary to make the amplitude of the plane wave exponentially small in the far past. This is achieved by adding a small positive imaginary part to the real
frequency $\omega$, $\omega \to \omega +i0$. 

\begin{figure}
\centering
\includegraphics[width=6cm]{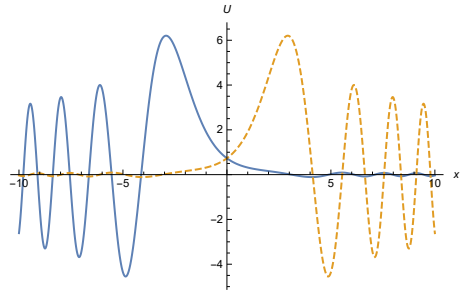}
\caption{Fundamental solutions of the parabolic cylinder differential equation (\ref{pcy}). The solution  $U(x)$ (solid line) falls exponentially, the solution $ U(-x)$ (dashed line) grows exponentially in the region $-2<z<2$. Here $C=1$.} 
\end{figure}

Let us to establish a connection between solutions of the equation (\ref{pcy}) on both sides of particular points $ z = -2C^{1/2} $
and $ z = 2C^{1/2} $. Since the transmitted wave exists in the region $ z> 2C^{1/2} $, we analyze the solution of the equation (\ref{pcy}) near the point $ z = 2C^{1/2} $.
Introducing the variable $ u = C^{1/6}(z-2C^{1/2}) $, we obtain the Airy equation,
$$
\frac{\partial^2\Sigma'}{\partial u^2}+u\Sigma'=0.
$$
We need to take a wave propagating in the positive direction, so we choose the Hankel function of the first kind with index 1/3,
\begin{equation}\label{airy}	
\Sigma'=u^{1/2}H^{(1)}_{1/3}\left(\frac{2}{3}u^{3/2}\right), \, u>0.
\end{equation}
The point $ u = 0 $ is a branch point, and the continuation of the solution of the (\ref{airy}) onto the region $ u <0 $ depends on how we pass around the point $ u = 0 $. In the upper half-plane of the
complex $ u $, $ u = \exp(-i\pi)|u| $, or in the lower half-plane, $ u = \exp(i\pi)|u| $. In the first case, the function $ H^{(1)}_{1/3} (2u^{3/2} / 3) $ will become the MacDonald function $ K_{1/3} (2|u|^{3/2}/3), \, u <0 $, and in the second case - into the modified Bessel function 
$ I_{1/3} (2|u|^{3/2}/3), \, u <0 $. Since according to the causality principle,
on which we have already discussed above, the pole lies in the lower half-plane when $ d \Omega / dr <0 $. Then it should be passed above. And we get an exponentially growing solution. The transmitted wave amplifies, and this is the amplification of the acoustic wave passing through the resonance from left to right, from the region of rapid rotation to the region of the slower rotation. This is the Landau effect of reverse damping, the wave energy is drawn from the energy of disk rotation. In the opposite case, $ d \Omega/dr> 0 $, the pole lies on the top, the wave attenuates. This is the Landau damping.

Thus, the interpretation of the passage of acoustic waves through resonance as the subbarrier tunneling , which still exists until now, is not correct, it was based on incorrectly understanding mathematical formulas.

In addition, in later works (\cite{tsang1}, \cite{tsang2}), equations of the type (\ref{eq2}) taking into account small derivatives $ d(\ln(\Omega/\Sigma))/dr \simeq r^{-1} << \lambda^{-1} $ were analyzed by authors to identify the effects of superreflection and instability of acoustic waves during their passage through the resonant layer. Small increments were found depending on the magnitude and sign of the derivative $d(\ln(\Omega/\Sigma)/dr $. However, for some reason not specified by the authors, solutions in the domains $ r_{-1} <r <r_{+1} $ and $ r<r_{-1} $ were obtained by continuing solutions from the domains
$ r>r_{+1} $ and $ r_{-1} <r <r_{+1} $, respectively, replacing $r-r_{+ 1} (r>r _{+1}) $ by $(r_{+1}-r)\exp (i\pi) (r <r_{+1}) $ and replacing $ r-r_{-1} (r> r_{-1}) $ by $ (r_ {-1}-r)\exp(i\pi) (r <r_{- 1}) $, respectively (see formulas (26,30,31) of \cite{tsang1}). This corresponds to a bypath of the pole in lower half-plane, which in our case contradicts the Landau's rule of bypath of a pole for $ d\Omega/dr <0 $.

\section{Conclusions}

We have shown that the collisionless dissipation of the angular momentum of a disk leads to the accretion of matter
onto the center. This requires the existence of a turbulent flow of acoustic waves propagating from the internal
edge of the disc outwards. Passing through the resonance regions, $ \omega = (n \pm 1) \Omega $, waves experience the exponential amplification
due to the reverse Landau damping. This results in a strong turbulence of the disk and in appearance of a wide
spectrum of acoustic turbulence. 

We emphasize that turbulence in a disk does not arise simply because of an instability, i.e. growth
of an initial disturbances in time, but by the passage of an acoustic wave through the resonant layer. Here collisionless inverse Landau damping is not temporary, but has the spatial character -
exponential growth of the wave amplitude in the radial direction.

An analogue of the parameter $ \alpha $ here is the turbulence power $ W $ (\ref{alpha}), which
has a quite definite physical meaning. The value of $ W $ can be calculated for a concrete
system in which an accretion disk is formed. We can also connect variations of radiation in the X-ray and the optical ranges ${\dot M} (\omega)$ with properties of an accretion disk.

Finally, it should be noted that in the presence of a magnetic field in an ionized disk, when the Alfv\'en velocity exceeds the acoustic velocity, it is necessary to consider a magnetosonic turbulence. However, the resonance in the region $ r_{-1} <r <r_{+1} $ will be of the same nature, since it is associated only with the neighbourship of the frequency of the wave $ \omega $ to the angular velocity of the disk rotation $ \Omega (r) $. Specifically, what happens with a magnetosonic turbulence in a magnetized disk requires special consideration.

\section{Acknowledgments}
This work was supported by Russian Foundation for Basic Research, grant number 17-02-00788.

\end{document}